\documentclass[twoside]{ilcws07}
\usepackage[latin1]{inputenc}
\usepackage[dvips]{graphicx,epsfig,color}
\usepackage{wrapfig,rotating}
\usepackage{amssymb,amsmath,array,gensymb}

\begin{document}
\title{
CALICE Si-W EM Calorimeter: Preliminary Results of the Testbeams 2006 
}

\author{C.~C\^arloganu$^1$ and A.-M. Magnan$^2$ on behalf of the CALICE Collaboration
\vspace{.3cm}\\
1- LPC Clermont-Ferrand, IN2P3/CNRS, UBP, France\\
2- Imperial College London, United Kingdom
}

\maketitle

\begin{abstract}
The CALICE~\cite{calice} Si-W electromagnetic calorimeter~\cite{ECAL} has been tested with electron beams (1 to 6 GeV) at DESY in May 2006, as well as electrons (6 to 45 GeV) and hadrons (6 to 80 GeV) at CERN in August and October 2006. Several millions of events have been taken at different incident angles  (from 0$^{\circ}$ to 45$^{\circ}$) and three beam impact positions. The ECAL calibration  is performed with muon beams and shows a good uniformity for nearly all channels. The large statistics available allows not only to characterise the ECAL physics performance, but also to identify subtle hardware effects.
\end{abstract}

\section{Introduction: the Calice ECAL Prototype}

The Si-W ECAL physics prototype is composed of 30 layers of $3\times 3$ wafers, each wafer having an array of $6\times 6$ pixels of $1\times 1$~cm$^{2}$. The two top rows of wafers are completed for the full depth in July 2006. 
The mechanical structure consists of tungsten sheets wrapped in carbon fibre, providing 15 alveola where slabs are inserted. One slab is made of two PCBs on each side of a tungsten layer, with the wafers conductively glued to the PCB. The very front end  electronics (VFE) provide preamplification and are located outside the active area, but mounted on the same PCB as the silicon wafers. The prototype is built of  three stacks, each of ten layers of alternating tungsten and silicon. Each stack has a different tungsten thickness: 1.4 mm or $0.4X_0$ per layer in the first stack, 2.8 mm or $0.8X_0$ per layer in the second stack and 4.2 mm or $1.2X_0$ per layer in the rear stack. This choice should ensure
\begin{wrapfigure}{r}{0.4\columnwidth}
\centerline{\includegraphics[width=0.35\columnwidth]{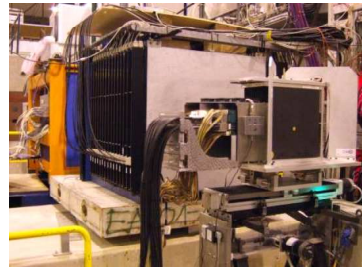}}
\caption{Testbeam setup at CERN, in August 2006. The ECAL is in front, followed by the analogue HCAL and TCMT.}\label{Fig:setup}
\end{wrapfigure}
a good resolution at low energy, due to the thin tungsten in the first stack, combined with a good containment of the electromagnetic showers, with an overall thickness of about 20~cm or $24X_0$ at normal incidence. To rotate to angles of $10^{\circ}$, $20^{\circ}$, $30^{\circ}$ and $45^{\circ}$ with respect to the beam axis, the three stacks, mechanically separate, are also translated laterally so that the beam still passes through all of them.\\
\indent The purpose of the testbeam phase is to validate the simulation against a realistic detector, as well as allowing to detect the potential hardware problems. Once the simulation is trusted, full detector  studies will lead to the optimisation of the calorimeter for a Particle Flow approach. The test setup at CERN is presented on Figure~\ref{Fig:setup}, and is simulated using Mokka~\cite{Mokka}.\\
\indent Three drift chambers are used for the tracking at CERN  and four at DESY. The ECAL, HCAL~\cite{HCAL} and TCMT~\cite{TCMT} follow, with the ECAL mounted on a movable stage for angle scans.

\section{Summary of the data taken}

Eight million triggers were taken at DESY during 14 days in May 2006, for seven beam energies (1, 1.5, 2, 3, 4, 5, 6~GeV), five angles (0$^{\circ}$, 10$^{\circ}$, 20$^{\circ}$, 30$^{\circ}$, 45$^{\circ}$) and three positions of the beam on the ECAL front face (center, border and corner of a wafer), with a minimum of 200,000 events per configuration. Six layers were not instrumented: the last eight layers had one dummy slab, one instrumented slab, and two dummy slabs.\\
\indent The August 2006, CERN beams allowed to take another 8.6 million triggers in ECAL only mode, with the full depth instrumented, for six beam energies between 6 and 45 GeV, and four angles. Pions were taken between 30 and 80 GeV in combination with the HCAL and TCMT. For calibration purposes, 30 million muon events were also taken parasitically to an experiment upstream.\\
\indent The setup in October 2006 was slightly different, with ECAL and HCAL at 6 cm from each other. 3.8 million triggers were taken with electrons and positrons from 6 to 45 GeV, and 22 million with pions from 6 to 80 GeV. Another 40 million muon events were added for calibration purposes.

\section{ECAL Calibration}

\subsection{Gain Calibration}
\label{gaincalib}

The current calibration of the ECAL prototype is  performed by using a set of $74$ high-statistics muon runs ( $\sim 250,000$ events each), taken during October $2006$ with another experiment upstream, providing a wide spread of the muon beam over the front face of the prototype.  The events are triggered with a 1 m$^2$ scintillator counter.\\
\begin{wrapfigure}{r}{0.45\columnwidth}
 \centerline{\includegraphics[width=0.4\columnwidth]{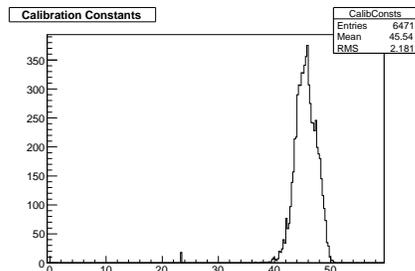}}
 \caption{Equivalent ADC value for the energy deposit of 1 MIP for the live cells of the ECAL prototype.}
\label{Fig:calib}
\end{wrapfigure}
\indent After pedestal substraction (see Section~\ref{peds}), the runs are reconstructed using a fixed global noise cut of half a Minimum Ionising Particle (MIP), 1~MIP being estimated to 50~ADC counts by former studies. To reject any remaining noise hits, it is required that the hits of one event form tracks characteristic of a MIP.\\
\indent The distributions obtained by channel are described by a convolution of a Gaussian and a Landau function. The calibration constant is defined as the most probable value (MPV) of the Landau function, while the standard deviation of the Gaussian defines the noise value for each cell. For 6403 out of the 6480 channels of the prototype, a calibration constant can be obtained via this method without further investigations. The remaining cells show a noise value that is unusually high, and are  treated by estimating the additional noise contribution, or by applying the calibration constant from one of their neighbours.\\
\indent An entire wafer (36 channels) seems to not be fully depleted at the applied voltage of 200V, resulting in a MIP peak at half the normal value. A relative value between the mean MIP signals of the wafer and its neighbours can be estimated, allowing a relative calibration of the cells.
0.14 \% of the channels give no output and were considered as dead. The calibration constants for all calibrated channels are histogrammed in Figure~\ref{Fig:calib}. The distribution is narrow, with almost all pads in the range 40 to 50~ADC~counts per MIP.  The small peak at 23.5 corresponds to the single incompletely depleted wafer.

\subsection{Pedestal}
\label{peds}

For all beam tests performed, the data acquisition consists of a fixed sequence of 500 pedestal events, 500 events with charge injection via the calibration chips, and then 20,000 beam events. The pedestal events are used to make a first estimate of the pedestal (mean value) and noise (RMS) per channel.\\
\indent It has been observed that the pedestals are not necessarily stable, but subject to a random shift affecting all channels of one layer with the same drift. This effect concerns several particular PCBs, is time dependant, and is attributed to the instabilities of the power supplies giving the pedestal lines, which are not isolated. To correct for these instabilities, the pedestals are recalculated on an event by event basis, by discarding all cells recording a signal, and iterating until the RMS of the distribution obtained with the remaining channels is of the order of the expected noise.

\subsection{Noise}
\begin{figure}[htbp!]
\begin{minipage}{0.32\columnwidth}
\centerline{\includegraphics[angle=-90,width=0.9\columnwidth]{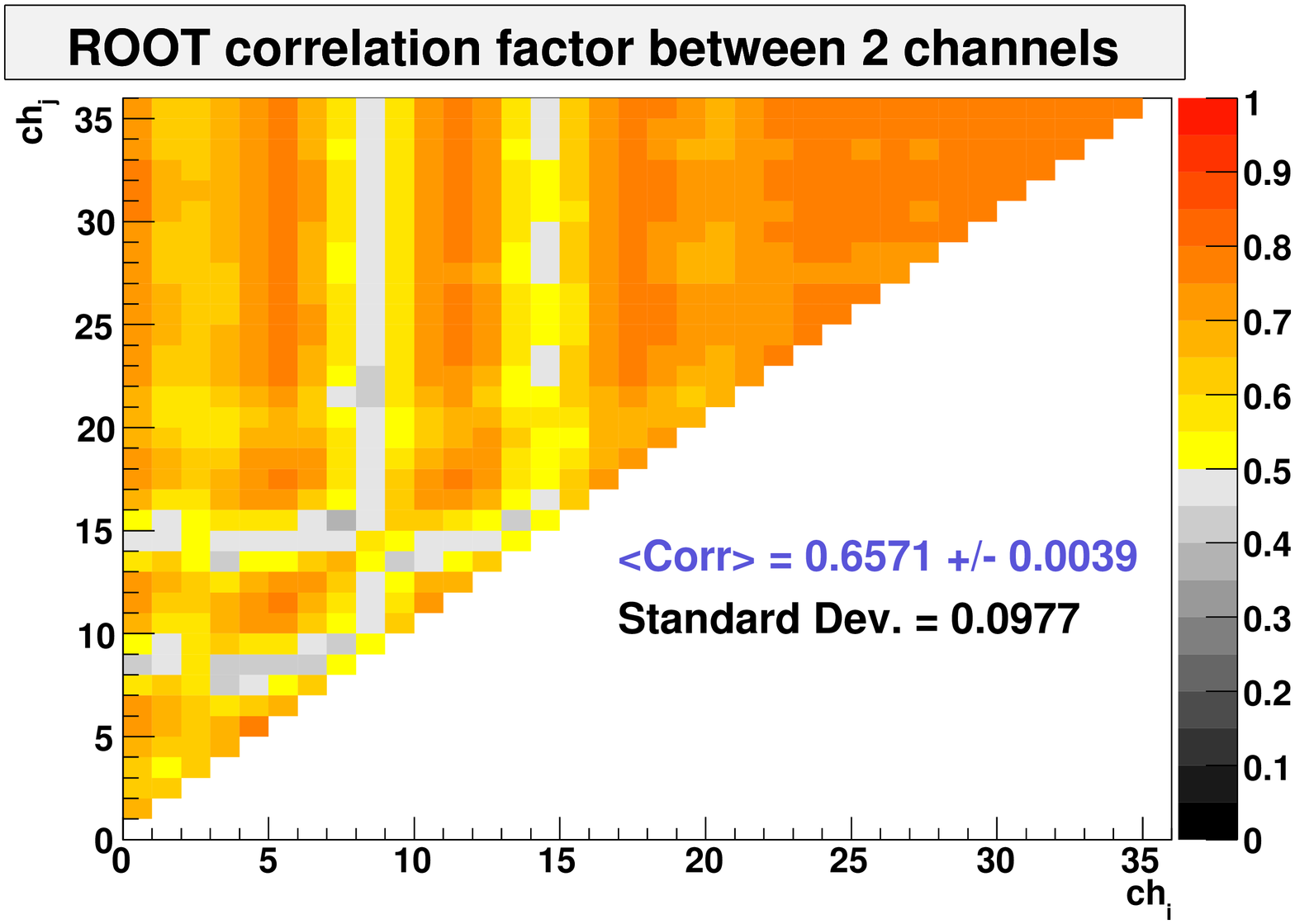}}
\end{minipage}
\hfill
\begin{minipage}{0.32\columnwidth}
\centerline{\includegraphics[angle=-90,width=0.9\columnwidth]{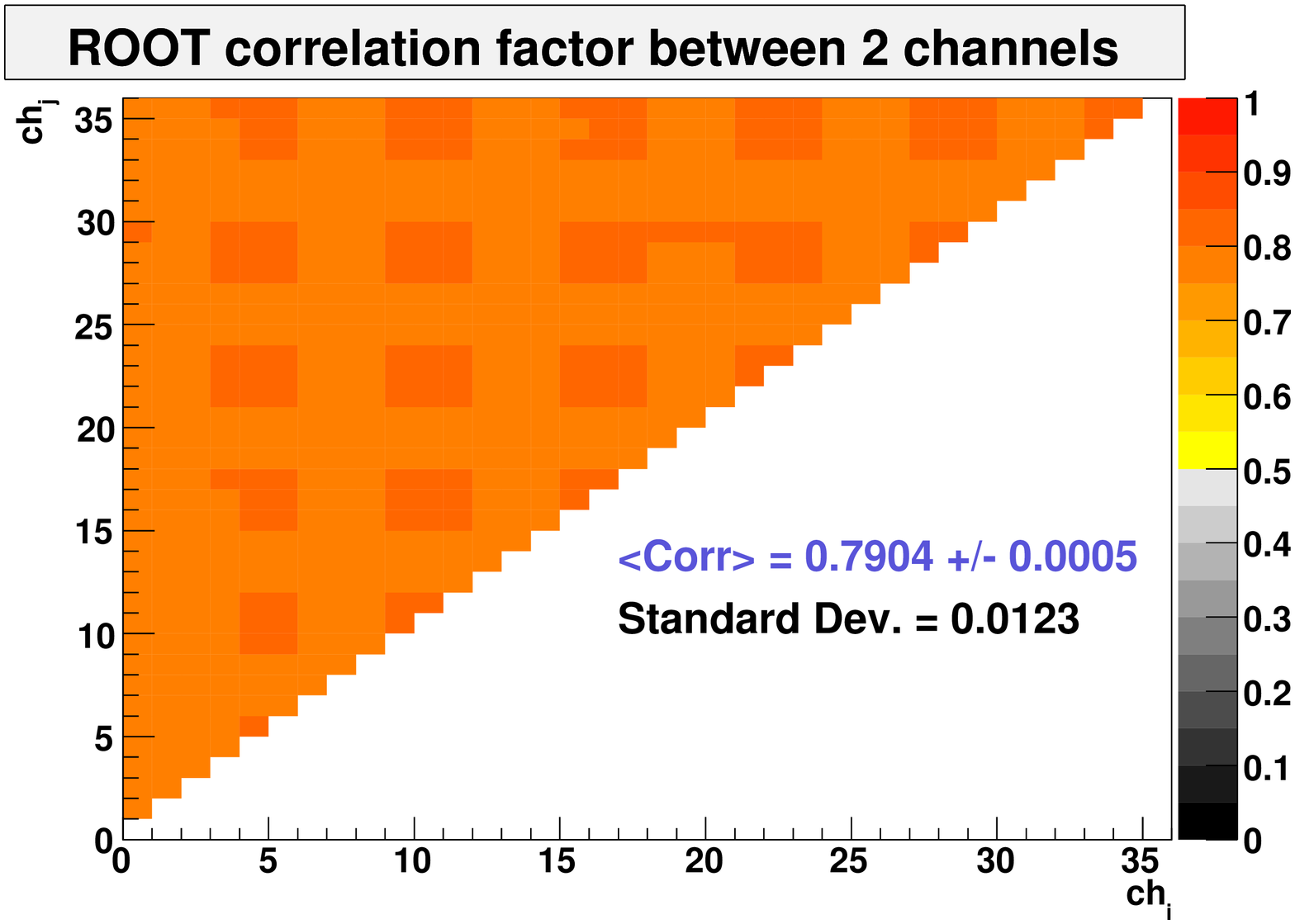}}
\end{minipage}
\hfill
\begin{minipage}{0.32\columnwidth}
\centerline{\includegraphics[angle=-90,width=0.9\columnwidth]{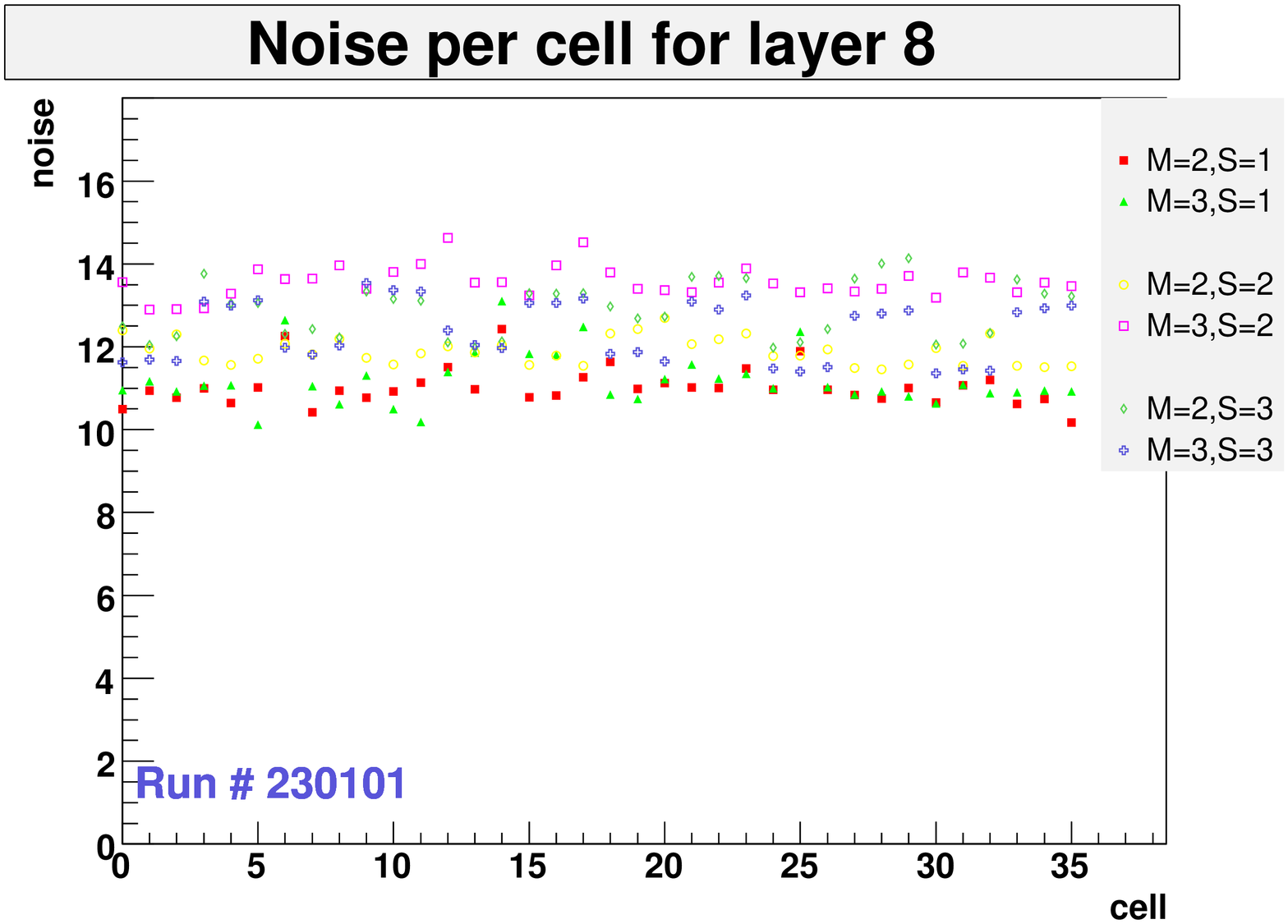}}
\end{minipage}
\hfill
\begin{minipage}{0.32\columnwidth}
\centerline{\includegraphics[angle=-90,width=0.9\columnwidth]{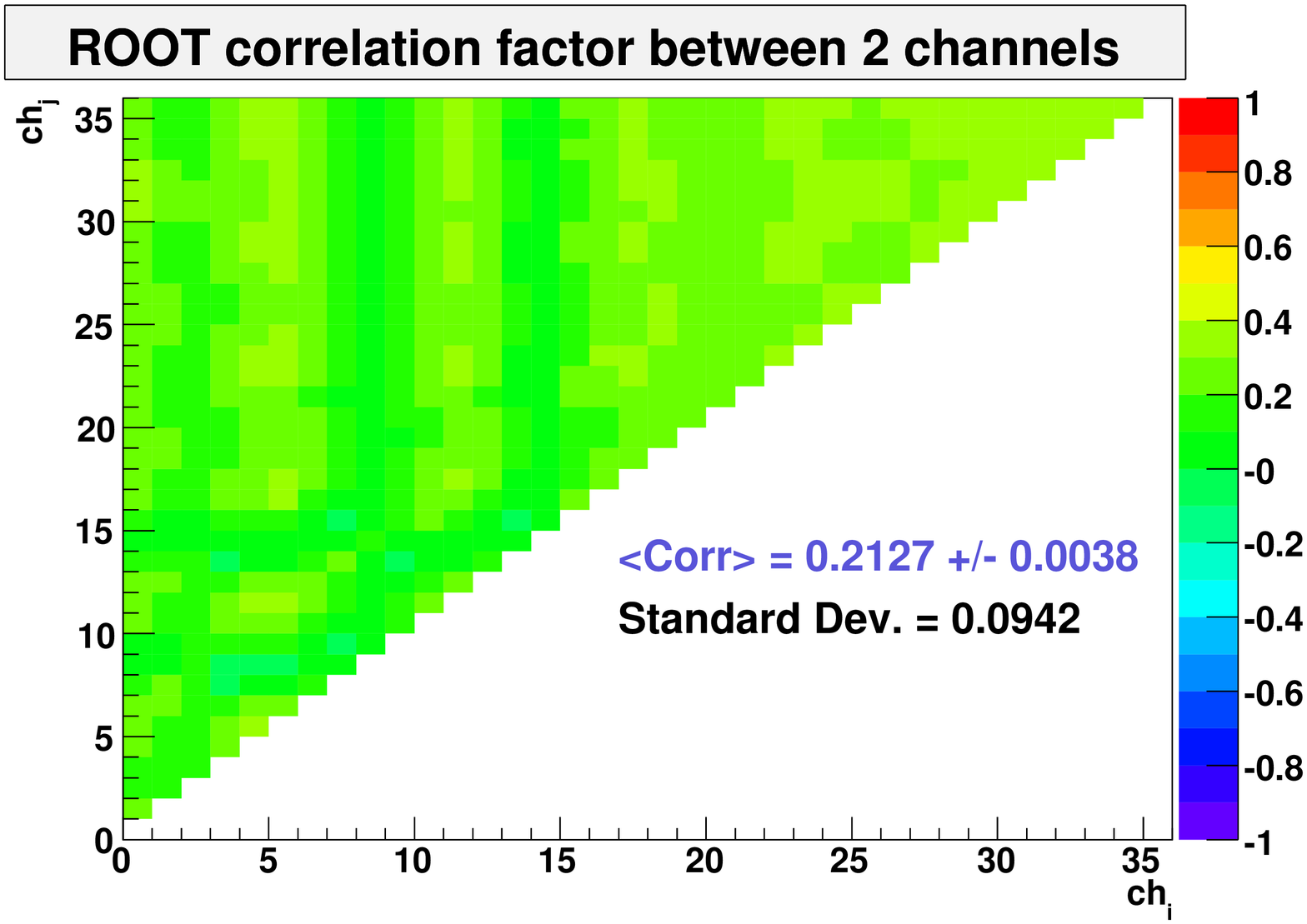}}
\end{minipage}
\hfill
\begin{minipage}{0.32\columnwidth}
\centerline{\includegraphics[angle=-90,width=0.9\columnwidth]{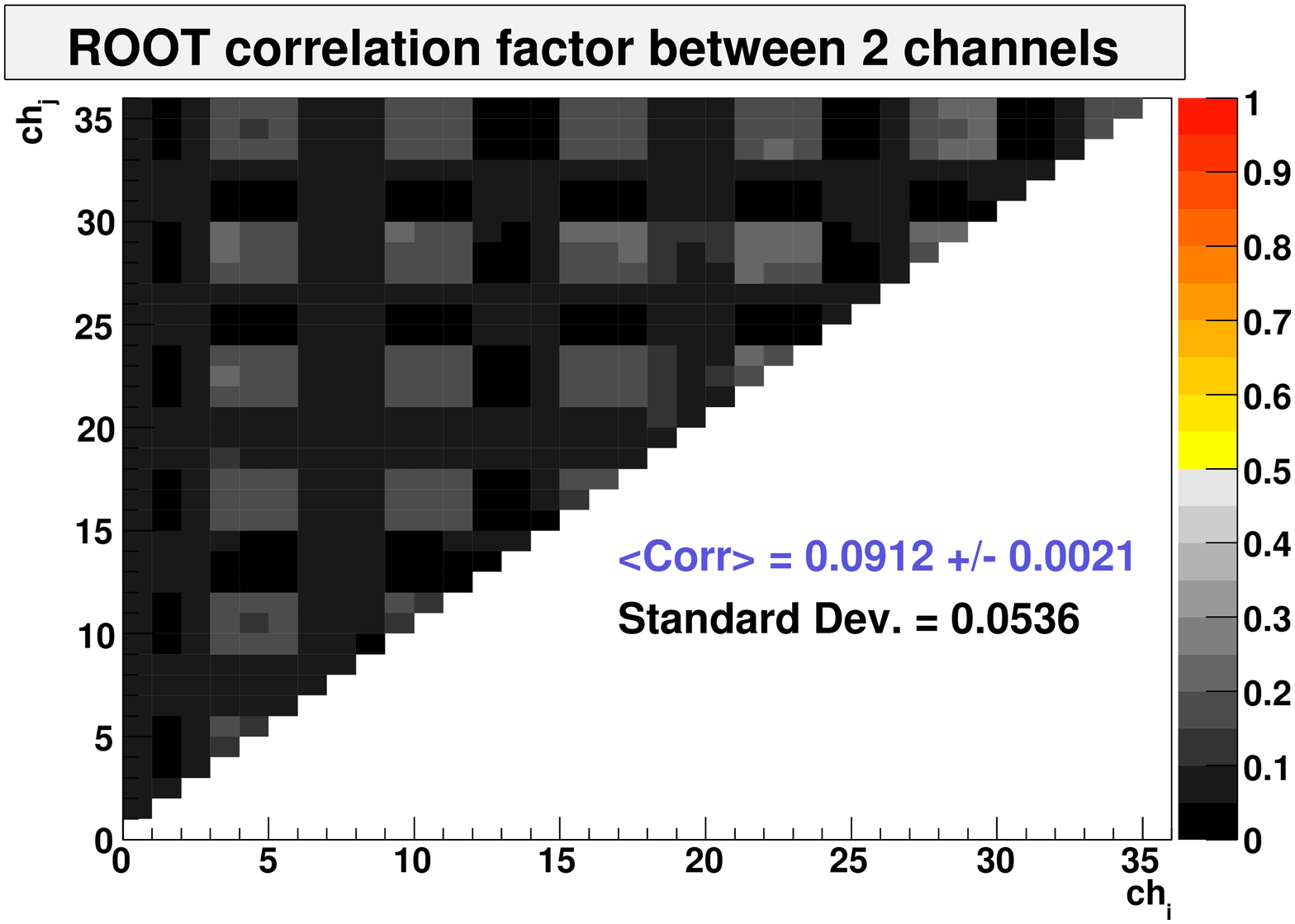}}
\end{minipage}
\hfill
\begin{minipage}{0.32\columnwidth}
\centerline{\includegraphics[angle=-90,width=0.9\columnwidth]{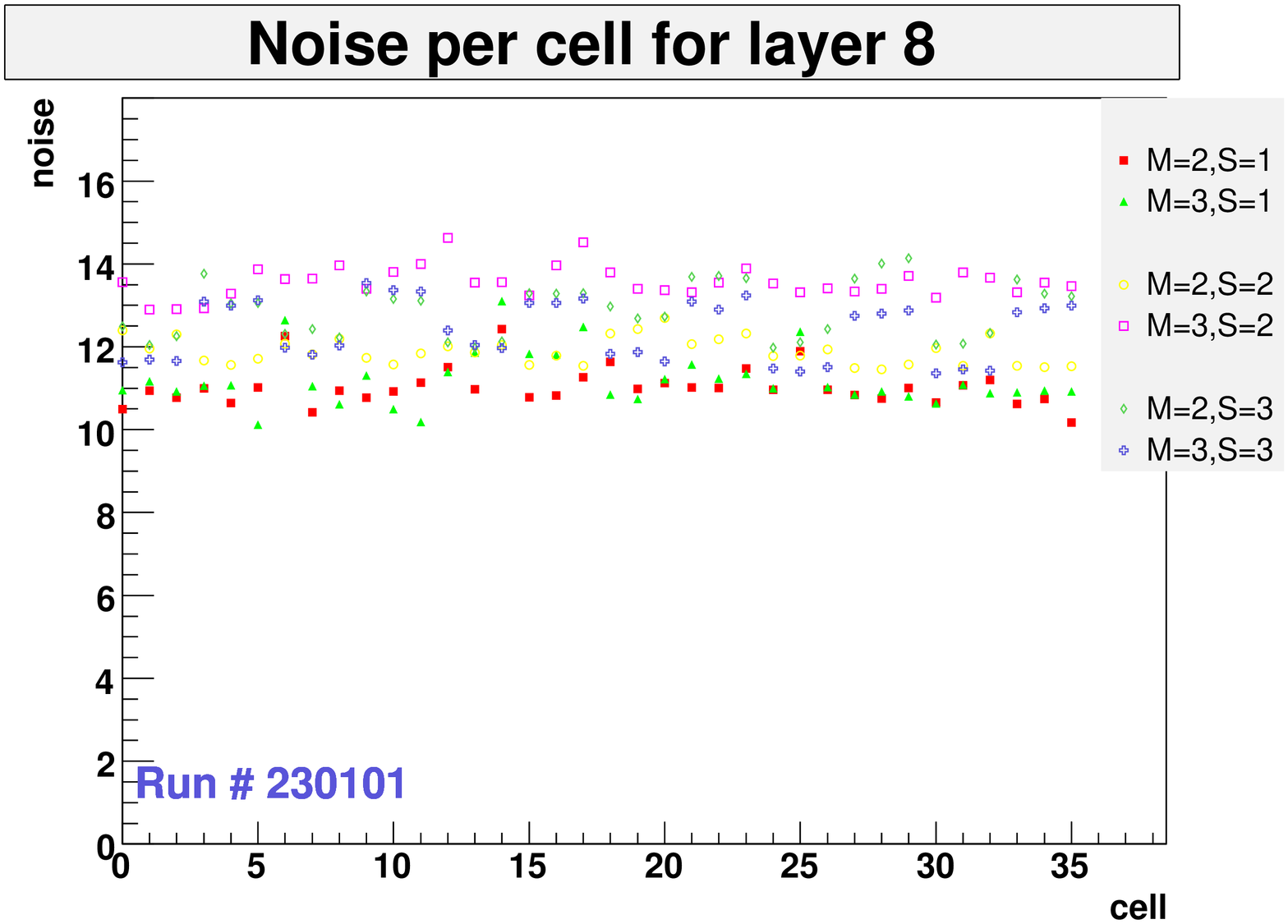}}
\end{minipage}
\caption{\label{Fig:noise}
Layer 8, {\it left:} module recording a signal, middle: neighbouring wafer, {\it right:} mean noise per wafer. {\it Upper row:} without corrections, and {\it bottom row:} with corrections.}
\end{figure}
In order to identify coherent noise, the correlation between pairs of channels is calculated in signal events. No clear correlation is seen in an entire PCB, except the one coming from the pedestal drift discussed above. The results are  thus shown only on a wafer basis. Figure~\ref{Fig:noise} displays the correlation factors (colour scale) as a function of the channel indices, numbered from 0 to 35, for a particular wafer affected by the pedestal drifts described above, and a run recorded at DESY with an energy of 6 GeV. Channels numbered 0 to 17 and 18 to 35 correspond to two different chips. The corresponding noise level per layer, for all wafers, is also presented in Figure \ref{Fig:noise}, on the right column.\\
\indent The results before any corrections are shown in the upper row of Figure \ref{Fig:noise}. The left plot represents the module in which the beam was directed. It can be seen that the region with signal shows less correlations, due to the fact that most pixels are discarded in the noise calculation. The middle plots show a neighbouring module, affected only by the global pedestal drift. The difference between these two wafers allows the identification of a crosstalk issue. All the channels of some wafers recording a high signal, like the one presented in Figure~\ref{Fig:noise} on the left, suffer from a pedestal shift towards negative values. This effect does not propagate to the neighbouring wafers. This seems to be random in space and in time, but it is clearly correlated with the intensity of the signal recorded. This effect is not yet understood, but under investigation. In order to correct for the induced correlated noise, the mean and standard deviation are calculated on an event-by-event basis, per wafer, after discarding the signal hits, and iterating over the channels taken into account in the sum. The bottow row on Figure~\ref{Fig:noise} shows the results after all the pedestal corrections described above. The corrections are performing well,  bringing the noise back to the normal level of 6 ADC counts. For this particular layer, affected by both problems (i.e. coherent noise on the PCB level due to an instability in the power supply, and crosstalk affecting the wafer recording a high signal), the wafer recording a signal still shows a remaining 20\% of correlations. The correlation is however completely removed for wafers affected only by the crosstalk problem, which confirms that the corrections are performing well.\\

\section{Background for physics performance studies: electron selection }
 \begin{wrapfigure}{r}{0.75\columnwidth}
 \vspace*{-0.5cm}
 \centerline{\includegraphics[width=0.75\columnwidth]{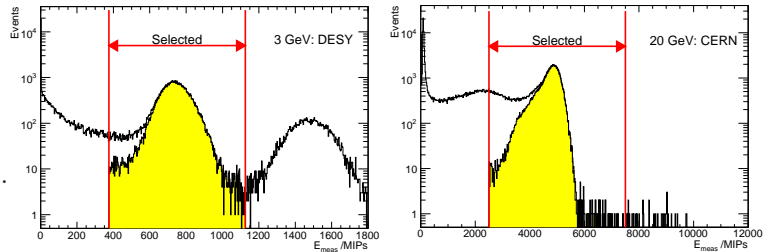}}
 \caption{Total ECAL energies for a 3~GeV e$^-$ DESY beam ({\it left}) and for 20~GeV at CERN ({\it right}), 
          with the energy selection windows. The shaded areas show the effect of the 
          cuts on the shower barycentre ({\it left}) and on the \v{C}erenkov counter signal ({\it right}).}
\label{fig:eselection}
\end{wrapfigure}

The selection of the single electron events is loose, in order to avoid bias:
\begin{itemize}
\item $E_{cell}\ge 0.6~\mbox{MIP}$ removes the noise.
    The threshold is about five times the average noise measured per cell.
\end{itemize}

\begin{itemize}
\item    the total energy recorded in the ECAL,
$ E_{\mathrm{raw}}$, should be in the range 
 $ 125 < \left(E_{\mathrm{raw}} (\mathrm{MIP})\right)/ \left(E_{\mathrm{beam}} (\mathrm{GeV})\right) < 375$. $E_{\mathrm{raw}}$  is 
computed with the three stacks weighted in the ratios 1:2:3, according to the 
tungsten thickness.
\end{itemize}
Further cuts are applied to some particular samples: the significant pion content of
some high energy electron runs at CERN is  reduced by using the 
threshold \v{C}erenkov counter, whereas the low energy halo coming with
some low energy DESY beams is rejected with additional cuts 
on the shower barycentre. The effect of these two additional cuts 
is indicated by the shaded regions in Figure~\ref{fig:eselection}.

\section{Energy Response, Linearity and Resolution}

The total response of ECAL is computed by summing the hit energies
in the three sections of the detector. If $E_1$, $E_2$, $E_3$ are the recorded energies in the first, second and third stack respectively, the total 
response is 
$E_{\mathrm{tot}}=(\alpha_1 E_1 +  \alpha_2 E_2 + \alpha_3 E_3)/\beta$.
The na\"ive choice for the weights $(\alpha_1,\alpha_2,\alpha_3)=(1,2,3)$ is generally used. It
reflects the relative 
thicknesses of the tungsten layers 
in each of the stacks, and hence the relative
sampling fractions. However, a weighting scheme  optimisation for energy resolution was performed as well, leading to the slightly different values of $(1.1,2.,2.7)$.
The normalisation $\beta$ has been arbitrarily fixed to  250~MIP/GeV.  

\begin{wrapfigure}{r}{0.7\columnwidth}
 \centerline{\includegraphics[width=0.7\columnwidth]{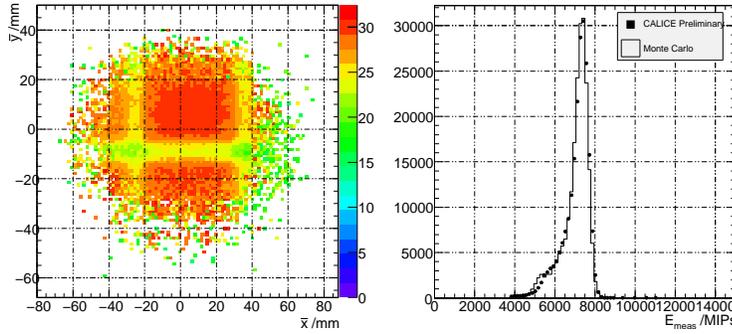}}
 \caption{{\it Left:} mean ECAL energy as a function of the shower barycentre $(\bar{x},\bar{y})$. {\it Right:}  total ECAL energy for a 30~GeV $\mathrm{e}^-$ beam, for data (points) and Monte Carlo (
   open histogram ).}
\label{fig:Edistrib}
\end{wrapfigure}

The guard rings create 2~mm non-active inter-wafer gaps, causing non-uniformities in the ECAL response,
 as illustrated in
Figure~\ref{fig:Edistrib}, left, where the mean energy is plotted 
as a function of the shower barycentre,
\mbox{$ (\bar{x},\bar{y})=\sum_i(E_ix_i,E_iy_i)/\sum_iE_i $}.
Dips in response are clearly visible 
and account for the asymmetric tail on the low side of the distribution of total energy (Figure~\ref{fig:Edistrib}, right), which is
reasonably well modelled by the simulation. The correction of these non-uniformities will be discussed in Section 6.

The beam profile, and thus the fraction of beam particles traversing the 
inter-wafer gaps, depends on the beam energy.  
Therefore, in order to avoid bias, a cut is applied on the shower barycentre position such as to select showers not affected by the gaps.

\begin{wrapfigure}{r}{0.7\columnwidth}
 \centerline{\includegraphics[width=0.7\columnwidth]{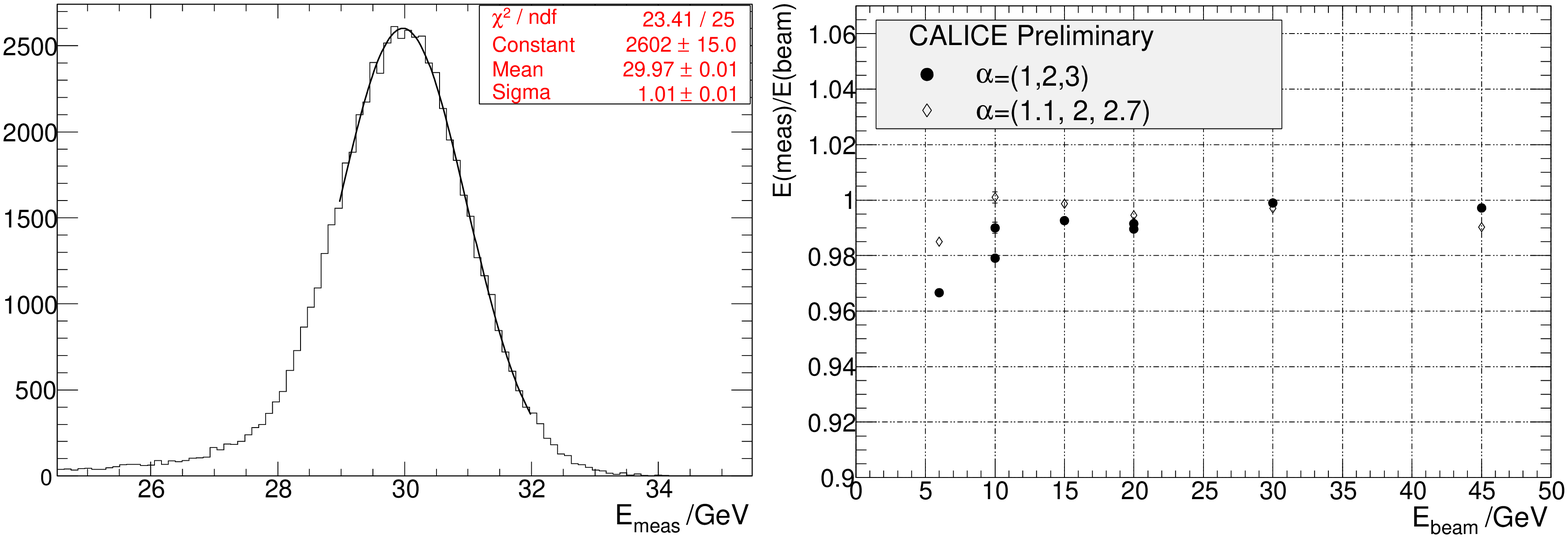}}
 \caption{{\it Left:} a  Gaussian fit to the measured energy, for  30~GeV e$^-$ data. 
{\it Right:}  ECAL energy response, divided by beam energy, as a function of beam energy.}
\label{fig:energy}
\end{wrapfigure}
To estimate the energy resolution and the mean calorimeter response, 
the distribution from  
Figure~\ref{fig:energy}, left, is fitted by a Gaussian in the asymmetric range of
$[-\sigma,+2\sigma]$ in order to 
reduce sensitivity to pion background and  to radiative effects upstream of the calorimeter, as well as  to any 
residual influence of the inter-wafer gaps. 
\begin{wrapfigure}{r}{0.45\columnwidth}
\vspace*{-1.cm}
 \centerline{\includegraphics[width=0.45\columnwidth]{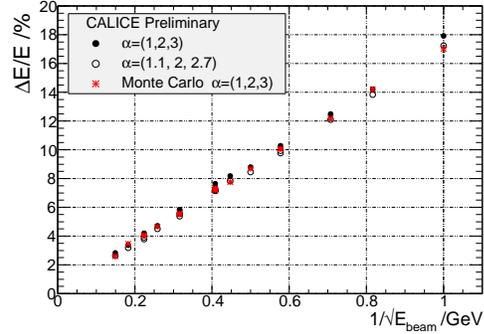}}
 \caption{ECAL energy resolution as function of $1/\sqrt{E}$. Two possible weightings of the ECAL stacks are compared.}
\label{fig:energy_res}
\end{wrapfigure}

The ratio of the reconstructed energy to the beam energy,  as function of the beam energy is shown in
Figure~\ref{fig:energy}, right, for the two choices of  weights. Non-linearities are at the $\%$ level. The linearity is somewhat better for the optimised weights.

The energy resolution, similar for both weightings,  is shown in Figure~\ref{fig:energy_res}. The Monte Carlo prediction, with
the na\"{i}ve weights, also shown, is in reasonably good agreement with the data.
The  resolution can be parametrised, for
the  na\"ive choice of weights and, respectively, the optimised one, as\\

\begin{displaymath}
  \frac{\Delta E}{E}(\%)= \frac{17.7\pm0.1}{\sqrt{E(\mathrm{GeV})} }
\oplus 1.1\pm0.1
\;\;\;\;\;\;\;\;\;\;\;
\frac{\Delta E}{E}(\%)= \frac{17.1\pm0.1}{\sqrt{E(\mathrm{GeV})} }
\oplus 0.5\pm0.2,
\end{displaymath}

\section{Interwafer gap corrections}
\label{Gapcor}
\begin{wrapfigure}{r}{0.7\columnwidth}
\vspace*{-0.5cm}
 \centerline{\includegraphics[width=0.7\columnwidth]{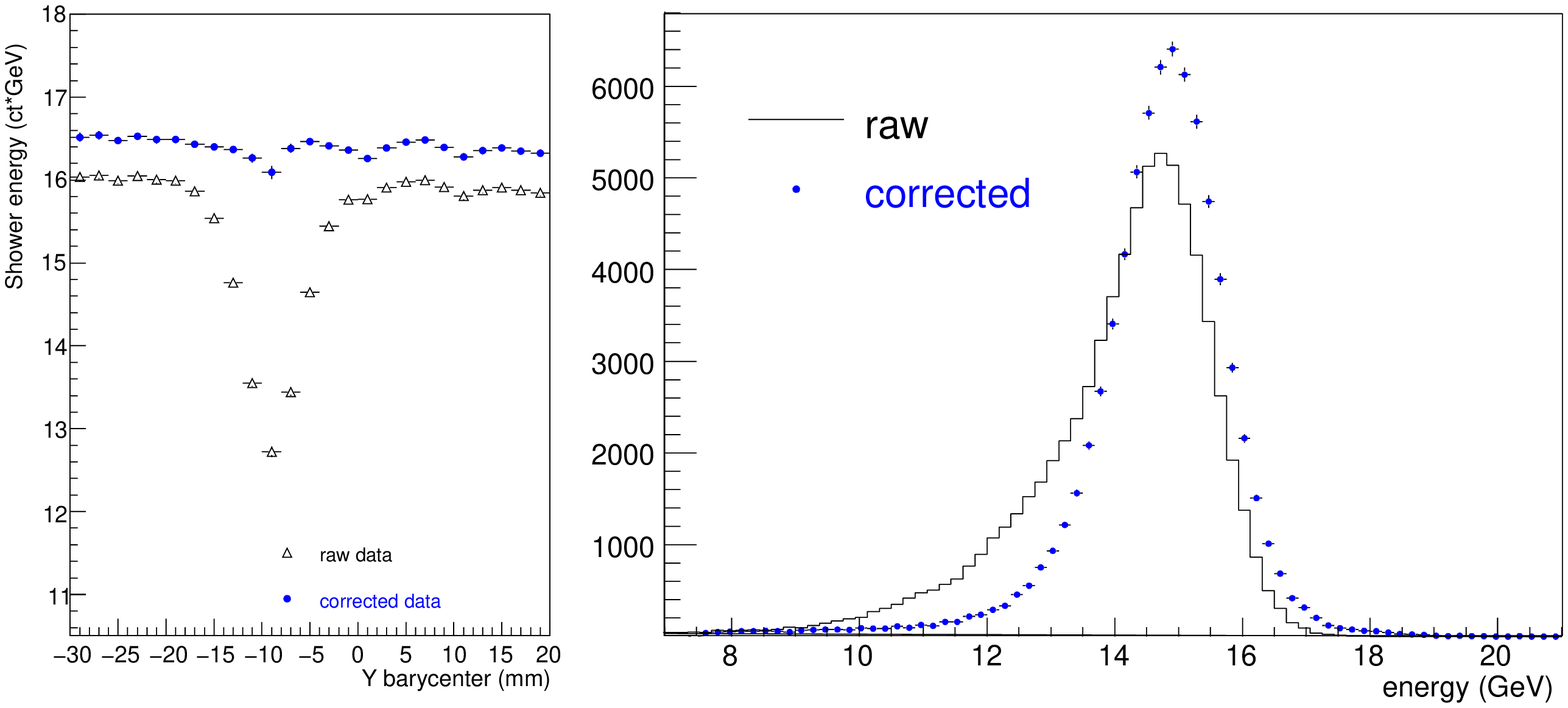}}
 \caption{15~GeV e$^-$, before  (black triangles) and after corrections (blue circles): total energy as function of the shower barycentre $\bar{y}$ ({\it left}), and energy distribution ({\it right}).}
\label{fig:gaps}
\end{wrapfigure}
The method used for correcting the interwafer gaps operates  at the event level and relies only 
on the calorimeter information: it parametrises the mean calorimeter response as function of the 
shower position in the calorimeter and applies subsequently corrective factors for each event according to this parametrisation. It is only geometrical, independent in $\bar{x}$ and $\bar{y}$ and without any explicit dependence on the shower energy.

The impact of the corrections is clearly illustrated in Figure~\ref{fig:gaps}, left, where the $y$ scan of ECAL is shown for the raw and, respectively the corrected data. The low energy tail  of the energy distribution is greatly reduced (Figure~\ref{fig:gaps}, right).
The resolution loss when going from the out of gap events to all the events (with corrections applied on the energy) is of the order of 10\%.  The corrections do not degrade the linearity.

When tracking information is available, it is possible to precisely calculate the shower position within each wafer. Subsequently, the ratio of the active to non-active areas crossed by the shower according to a mean shower shape can be estimated and used to correct the energy recorded in each layer. 

\section{Shower development}

\begin{figure}[h!]
\begin{minipage}{0.48\columnwidth}
 \centerline{\includegraphics[width=0.95\columnwidth]{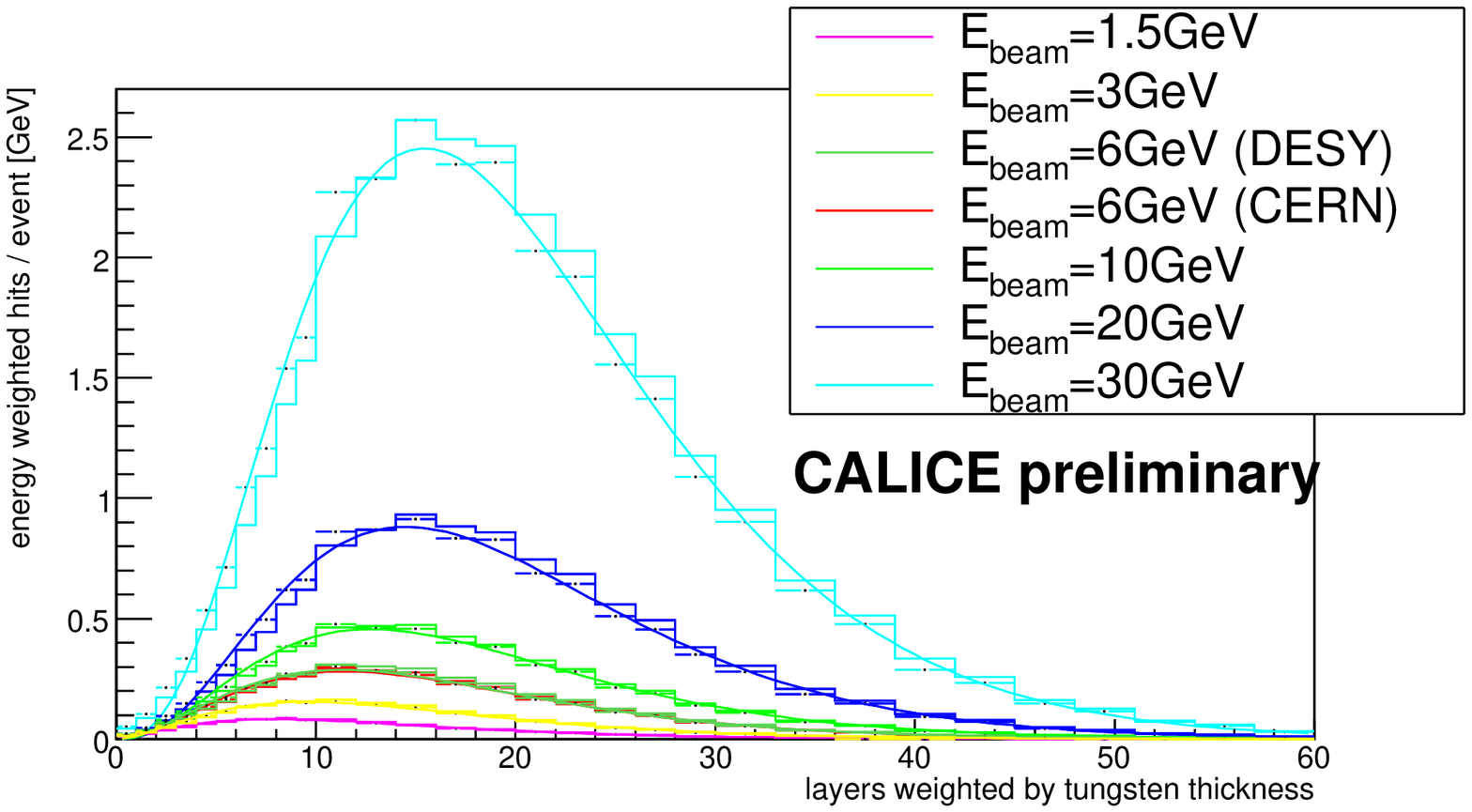}}
\caption{Longitudinal shower profile for the data 
    (points with statistical uncertainties) and Monte Carlo
    simulation (histogram). The smooth curve is the used parametrisation of the
  shower profile.}
\label{fig:long_dev}
\end{minipage}
\hfill
\begin{minipage}{0.48\columnwidth}
\vspace*{-1cm}
 \centerline{\includegraphics[width=0.95\columnwidth]{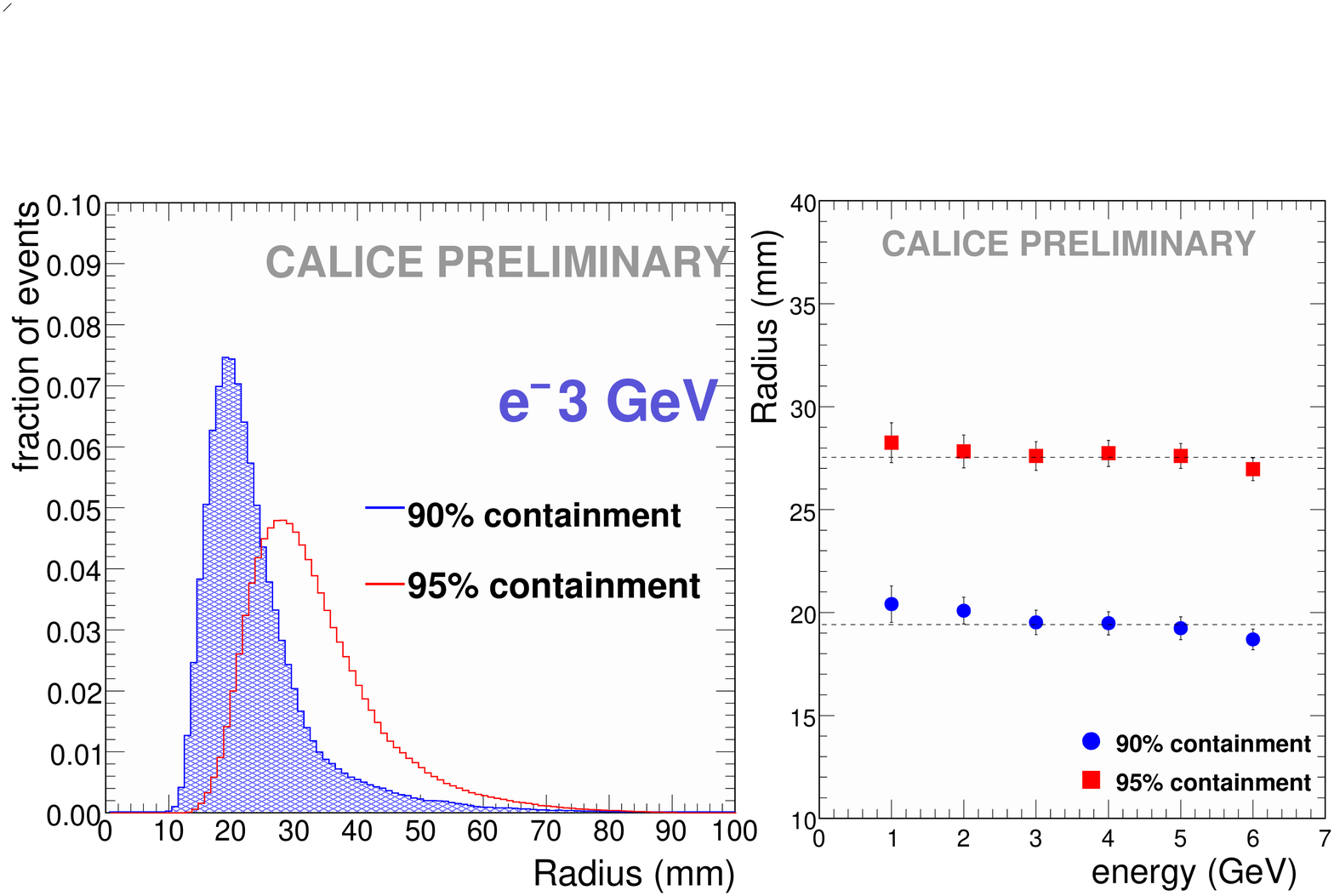}}
 \caption{{\it Left:} distribution of radii for 90\% and 95\% signal containment (3~GeV e$^-$, normal incidence, centre of wafers). {\it Right:}  radii for 90\% and 95\% signal containment of e$^-$ as function of the beam energy.}
\label{fig:rad_dev}
\end{minipage}
\end{figure}

Only events outside the interwafer gaps were used to characterise the longitudinal development of the shower.
The mean energy distribution is well fitted by the standard parametrisation,
$\gamma(t)=c \,t^{\alpha} \exp(-\beta t)$,
 where $t$ is the calorimeter depth, $c$ is an overall normalisation,
 $\alpha$ and $\beta$ are constants (Figure~\ref{fig:long_dev}).
 The position of the shower maximum grows logarithmically with the beam energy. 

An important issue in the development of a calorimeter is 
to achieve the smallest possible effective Moli\`ere radius, in order to 
provide the best  shower separation. It requires the use 
of an absorber with a small intrinsic  
Moli\`ere radius~($R_M$), but also the minimisation of the gaps
between the absorber layers.  

Figure~\ref{fig:rad_dev}, left shows the event distribution  for 90\% and 95\% levels of 
signal containment with respect to the radius. 
The results for the various energies studied 
are summarised in Figure~\ref{fig:rad_dev}, right. The points correspond to the peak position of each radius distribution.
At 90\% (95\%) shower containment the corresponding radius, often quoted as 1~$R_M$,
is about 20 (28)~mm.

The geometry of the ECAL prototype, with  2.2~mm thick interlayer gaps leads to  
an effective  Moli\`ere radius  which is expected to be larger by 
a factor of about 2 with respect to $R_M$ of solid tungsten. 
The results from the test beam studies 
are therefore in agreement with expectations. 
R\&D effort towards the use of Si pads with integrated readout is 
under way and will hopefully lead to a significant decrease of the interlayer gap
and therefore of the ECAL effective Moli\`ere radius.

\section{Spatial and angular resolution of ECAL}
The spatial and angular resolution of the ECAL are studied with
the DESY data at normal incidence.
The shower direction and position at the ECAL front face are
constructed on an event-by-event basis using a linear two-parameter chi-square 
fit to the shower barycentre positions in each layer for the $x$ and $y$
coordinates separately. The correlation matrix is determined from 
simulations for each beam energy. 

\begin{wrapfigure}{r}{0.7\columnwidth}
 \centerline{\includegraphics[width=0.7\columnwidth]{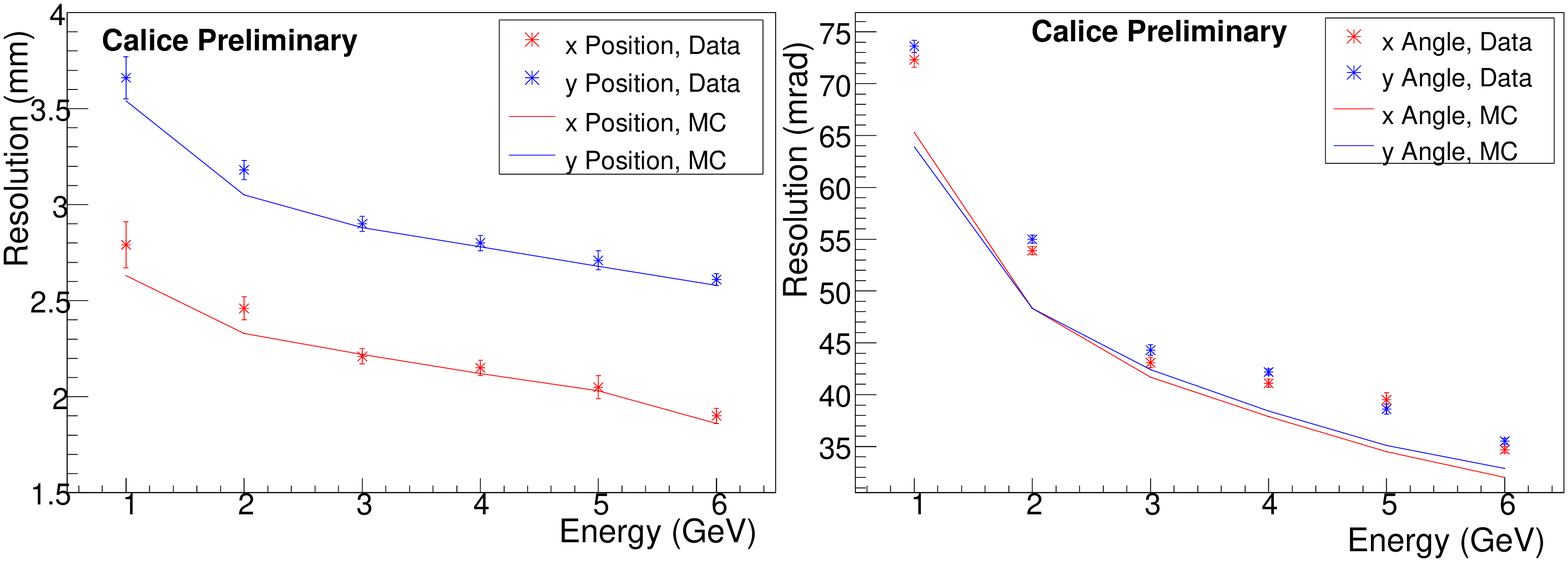}}
 \caption{  Resolutions in  position (left) and angle (right) as a function
            of the beam energy. The data are shown as points with error
            bars. The simulation expectation is shown by the continuous line.}
\label{fig:tracking}
\end{wrapfigure}
The fit results are compared
with the position and angle measured by the tracking system. The expected e$^-$ position and direction at ECAL front face
is obtained from  a linear fit of the drift chambers. Sources of systematic uncertainties as  residual misalignement, material modelling, and background rate are estimated for the extrapolation to the ECAL front face.

The ECAL resolution, deconvoluted from the tracking errors, is displayed in Figure~\ref{fig:tracking}.

\section{Conclusion}
The Si/W ECAL prototype was presented, as well as the results of the first beam tests at DESY and CERN during 2006. The prototype calorimeter was further exposed to beam at CERN in summer 2007. Analysis of the data collected is in progress.




\begin{footnotesize}




















%










\end{footnotesize}




\end{document}